\documentclass[aps,jcp,twocolumn,groupedaddress,showpacs]{revtex4} 
\usepackage{graphicx,amsmath}
\begin{document}

\title{Strain hardening in bidisperse polymer glasses: Separating the roles of chain orientation and interchain entanglement}
\author{Robert S. Hoy}
\email{robert.hoy@yale.edu}
\affiliation{Materials Research Laboratory, University of California, Santa Barbara, CA 93106}
\author{Mark O. Robbins}
\affiliation{Department of Physics and Astronomy, Johns Hopkins University, Baltimore, MD 21218}
\date{\today}
\begin{abstract}
The effects of entanglement and chain orientation on strain hardening in
glassy polymers are separated by examining mixtures of chains with
different lengths.
Simulations show that the
orientation of a molecule of a given chain length is the same in 
monodisperse systems and bidisperse mixtures,
even when entangled and unentangled chains are mixed.
In addition, the stress in mixtures is equal to the weighted average of
the stresses in monodisperse systems.
These results indicate that chains contribute independently to strain
hardening, that chain orientation is determined by local interactions
with the surrounding glass, and that entanglements play at most
an indirect role in strain hardening in the range of strains
typically studied.
We discuss these results in the context of recent microscopic theories.
\end{abstract}
\maketitle

\section{Introduction}

The mechanical performance and failure modes of glassy polymers are
strongly affected by strain hardening; an increase in the stress
needed to deform systems as the strain increases.
There has been great interest in understanding the
mechanisms of strain hardening so that it can be predicted and
optimized for applications.
Recent studies have challenged the traditional view that entanglements
between polymer chains control strain hardening, and suggest that
orientation of individual chains plays the dominant role.
This paper uses simulations of polymer mixtures to separate the
roles of entanglements and orientation.

The most widely used \cite{haward68,arruda93b} theories of
strain hardening are based on rubber elasticity.
They assume that entanglements between polymer chains act like
chemical crosslinks in a rubber and that
strain hardening arises from the decrease in entropy of 
an affinely deforming entanglement network \cite{foot3}.
The stress is written as
\begin{equation}
\sigma(\bar{\lambda}) \simeq \sigma_{flow} +  G_R g(\bar{\lambda})
\label{eq:sigma}
\end{equation}
where $\sigma_{flow}$ is the plastic flow stress,
$G_R$ the ``strain hardening modulus'',
$\bar{\lambda}$ the stretch tensor describing the macroscopic
deformation, and $g$ a dimensionless function that describes the
reduction in chain entropy.
This approach has had much success in fitting experimental stress strain
curves, but the fits seem inconsistent with the underlying
microscopic model \cite{kramer05}.
One difficulty is that $G_R$ is two orders
of magnitude larger than expected near the
glass transition temperature $T_g$.
More troubling is that
strain hardening grows with decreasing temperature,
while any entropic force should decrease.
A final difficulty is that the theory does not
explain the plastic flow stress, which must be
added by hand.

A number of recent experiments and simulations have examined the
origins of these difficulties
\cite{vanMelick03,kramer05,dupaix05,wendlandt05,hine07,casas08,govaert08,hoy06,hoy07,hoy08,lyulin04,robbins09,vorselaars09,vorselaars09b}.
One realization is that 
the two terms on the right hand side of Eq.\ (\ref{eq:sigma}) 
are of similar
magnitude and must arise from similar mechanisms.
Experiment, theory and simulations all show that  $\sigma_{flow}$ and $G_R$ are
linearly related when pressure, $T$ or strain rate is varied \cite{govaert08,chen09,vorselaars09,robbins09}.
This connection explains the large magnitude of $G_R$ and the
fact that it grows with decreasing temperature as is typical
for the plastic flow stress.
A direct connection between $\sigma_{flow}$ and $G_R$ has been
established in the athermal limit ($T \rightarrow 0$),
where both contributions to the stress are given by a single
scaling factor times the rate at which interchain bonds break during
local plastic rearrangements \cite{hoy07,hoy08}.

Simulations also show strain hardening occurs in polymers that are too
short to form the entangled network that is
assumed in rubber elasticity theories \cite{lyulin04,hoy07,hoy08,vorselaars09b}.
Moreover,
the strain hardening in these unentangled systems can be mapped to that of
entangled chains if the macroscopic deformation $\bar\lambda$ in
Eq. \ref{eq:sigma}
is replaced \cite{hoy07,hoy08} by an
effective stretch tensor $\bar{\lambda}_{eff}$ that describes the
orientation of individual chains:
\begin{equation}
\sigma(\bar\lambda) = \sigma_{flow} + G_R^0 g(\bar\lambda_{eff}),
\label{eq:sigmaeff}
\end{equation}
where $G_R^{0}$ is the value of $G_R$ in the long-chain limit.
The orientation of entangled chains is consistent with an affine 
deformation, $\bar{\lambda}_{eff}=\bar{\lambda}$.
For unentangled systems the orientation is consistent with
a subaffine deformation by $\bar{\lambda}_{eff}$.

The above findings support the notion
that entanglements play an indirect role and that
strain hardening is produced by the increased rate of local plastic
rearrangements that are needed to maintain chain connectivity
as chains orient \cite{argon73,hoy07,hoy08}.
In contrast, other recent experiments \cite{gsell97,vanMelick03}
suggest that entanglements play a central role.
They find that $G_R$ is directly
proportional to the entanglement density $\rho_e$ in both pure
systems and polymer mixtures.
However, there are other explanations for this correlation.
For example, an increase in chain stiffness increases $\rho_e$
and also stretches the chains so that more plastic rearrangements are
required to maintain chain connectivity
\cite{mckechnie93,fetters94,hoy08,vorselaars09b}.

In this paper we examine strain hardening of polymer mixtures.
Using polymers with different chain lengths but the same interactions,
stiffness, etc.,
allows us to isolate the roles of entanglements and orientation.
The simulation results show that the orientation of chains plays the dominant
role and that the response of mixtures has a surprisingly simple
mean-field form.
The variation of chain orientation with macroscopic stretch is
nearly independent of the length of surrounding chains.
In addition, the stress in mixtures is a simple weighted average
of results for pure systems.
These results suggest that strain hardening can be modeled as a sum
of contributions from individual chains as they are deformed by
interactions with the surrounding glass and produce local plastic
rearrangements in the glass to maintain their connectivity.

\section{Polymer Model and Methods}

Molecular dynamics (MD) simulations are performed using a coarse-grained
bead-spring polymer model \cite{kremer90}.       
All monomers have mass $m$ and interact via the truncated and shifted Lennard-Jones potential
\begin{equation}
U_{LJ}(r) = 4u_{0}[(a/r)^{12} - (a/r)^{6} - (a/r_{c})^{12} + (a/r_c)^{6}] \ \ ,
\end{equation}
where $r_{c} =1.5a$ is the cutoff radius and $U_{LJ}(r) = 0$ for $r > r_{c} $.
Covalent bonds between adjacent monomers on a chain are modeled using the
FENE potential
\begin{equation}
U(r) = -0.5kR_{0}^2 {\rm ln}(1 - (r/R_{0})^{2}) \ \ ,
\end{equation}
with the canonical \cite{kremer90} parameter choices
$R_{0} = 1.5a$ and $k = 30u_{0}/a^{2}$.   
The equilibrium covalent bond length is $l_0 = 0.96a$.
We consider semiflexible chains with an angular potential $U_{bend} = k_{bend}(1- cos(\theta))$, where $\theta$ is the angle between consecutive bond vectors along a chain.
All quantities are expressed in units of the molecular diameter $a$, binding energy $u_{0}$, and characteristic time $\tau_{LJ} = \sqrt{ma^{2}/u_{0}}$. 
Periodic boundary conditions are imposed, with periods $L_i$ along directions $i=x$, $y$, and $z$.

Simulations are performed using the same protocols as in Refs.\ \cite{hoy06,hoy07,hoy08}. 
A rapid quench is applied to well-equilibrated melts \cite{auhl03}, producing glasses at the desired temperature $T$.
The rapidity of the quench suppresses aging and strain softening.
The simulations presented here are performed at $T = 0.2u_0/k_B$ and $T = 0.275u_0/k_B$.
Both values are well below the glass transition temperature
$T_g \simeq u_0/3k_B$ \cite{rottler03c},
yet sufficiently high to observe significant thermally-activated
rate-dependent relaxation \cite{hoy06}.
Uniaxial compression is
performed at constant strain rate $\dot\epsilon \equiv \dot{L}_{z}/L_z$.
The volume stays nearly constant during compression.  
In this case the function describing the entropy loss
in Eq. \ref{eq:sigma} can be written
as $g(\lambda) = 1/\lambda^2 - \lambda$ where $\lambda=L_z/L_{z}^{0}$ is
the stretch relative to the initial size $L_{z0}$.

The strain rates used range from $10^{-6}/\tau_{LJ}$ to 
$10^{-4}/\tau_{LJ}$. 
While these rates are higher than experiment,
previous studies in this range of
$\dot\epsilon$ and $T$ have been shown to capture many aspects of experiments,
including logarithmic rate dependence \cite{rottler03c},
creep \cite{warren07}, and a linear
relationship between flow and hardening modulus
\cite{robbins09}.
Our rates are also orders of magnitude slower than those obtainable
with more computationally intensive united atom models
(e.\ g.\ Ref.\ \cite{vorselaars09b}).
At the higher rates studied with these potentials, unentangled chains
deform nearly affinely because there is insufficient time for
activated motion of chains relative to their neighbors.
This motion is particularly important for the chain length dependence of
strain hardening considered here.

Systems contain short and long chains with $N_{short}$ and $N_{long}$
monomers, respectively.
The total number of monomers is $N_{tot}$ and the
weight fraction of short chains is $f$.
We present results for $N_{long}=350$, $10 \leq N_{short} \leq 25$,
and $k_{bend}=0.75 \epsilon$,
but equivalent results were obtained for a range of other values.
For this $k_{bend}$, the entanglement length is $N_e \sim 40$
monomers \cite{everaers04}.
In experiments, unentangled glasses with $N \lesssim N_e$
typically exhibit brittle
fracture, but fracture is suppressed in simulations \cite{lyulin04,hoy06},
perhaps because
of the small system size and periodic boundary conditions.

\section{Results}

\begin{figure}[h]
\includegraphics[width=3.25in]{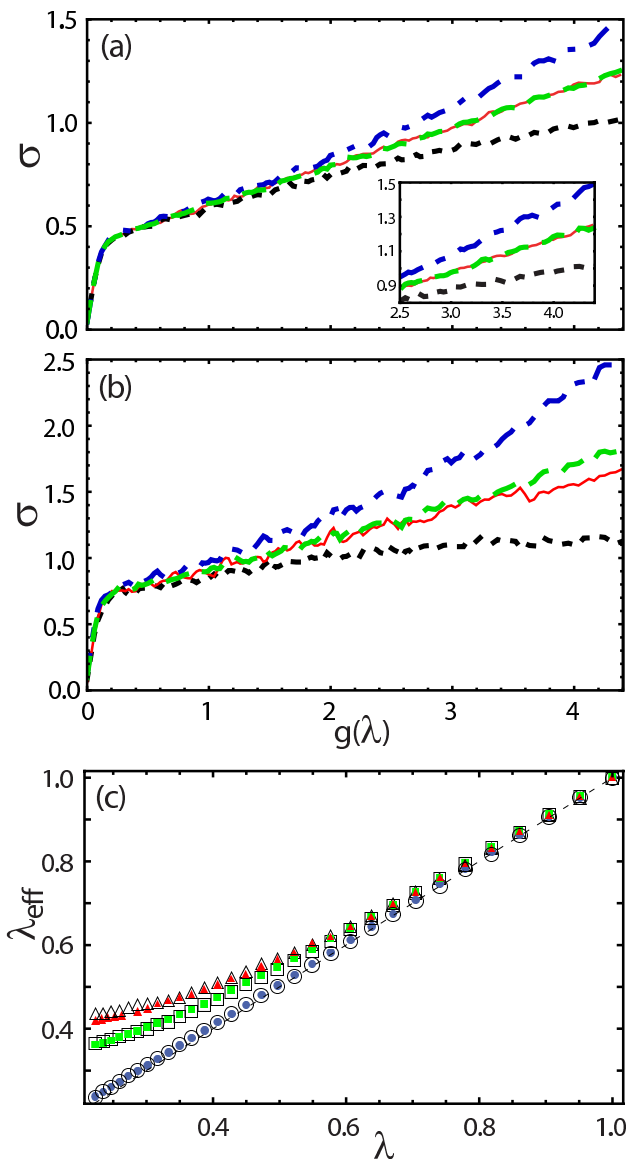}
\caption{Panels (a-b): Stress plotted against $g(\lambda)$ for uniaxial
compression of monodisperse glasses with length $N_{long}$ (dash-dotted)
or $N_{short}$ (dotted) and of 50/50 mixtures of the two lengths (dashed).
The solid curves show the average of the two monodisperse systems.
Here $k_{bend}=0.75$ ($N_e=39$).  
Panel (a): $N_{short}=25$, $T = .275u_0/k_B$ and
$|\dot\epsilon| = 10^{-5}/\tau_{LJ}$.
The inset shows a blowup of the data at large $g$ \cite{foot7}.  
Panel (b): $N_{short}=10$, $T = 0.2u_0/k_B$ and $|\dot\epsilon| = 10^{-4}/\tau_{LJ}$.
Panel (c): Variation of $\lambda_{eff}$ with $\lambda$ for pure systems with
$N=350$ (solid circles), $N=25$ (solid squares), and $N=10$ (solid triangles).
Open symbols of the same shape show results
for chains of each length in $f=0.5$ mixtures.
}
\label{fig:linsuperpos}
\end{figure}

Panels (a) and (b) of Figure \ref{fig:linsuperpos} show the stress as
a function of $g(\lambda)$ for mixtures with $N_{short}=10$ and 25,
respectively. 
Results for $f=0.5$ lie directly between the results for pure systems
with $f=0$ and 1.
Also shown is the weighted average of the pure system results
\begin{equation}
\sigma_{ave} = (1 - f)\sigma_{long} + f\sigma_{short}
\label{eq:sigmaavg}
\end{equation} 
for $f=0.5$.
Within our statistical accuracy,
the stress in the mixtures is equal to this weighted average.
We have verified that this ``stress superposition principle''
holds for different $k_{bend}$, chain lengths, temperatures and $f$.
The only discrepancies are observed
at large $g$ ($>3.5$) when
chains in entangled systems have stretched nearly taut and significant
energy is stored in the covalent bonds.
Experiments only reach such large stretches when special sample preparation procedures are used \cite{arruda93b,hasan93}.

Equation (\ref{eq:sigma}) would predict a linear increase in stress with $g$.
The systems in Fig.\ \ref{fig:linsuperpos} were chosen to exhibit
deviations from this straight line behavior, but
still obey stress superposition.
Deviations from linearity
in experiments are often fit by including
non-Gaussian (finite chain length) corrections to
the network entropy \cite{arruda93b}.
Previous simulations are not consistent with this interpretation
\cite{hoy07,hoy08}.
They show that the upturn in stress for entangled systems
reflects energy stored in the glass, while the downward curvature for
unentangled systems results from subaffine deformation of the chains
($|ln(\lambda_{eff}))| < |ln(\lambda)|$).
Despite the different curvatures for the long and short
chains in Fig.\ \ref{fig:linsuperpos} their contributions to the
stress in mixtures are simply additive.

These results do not conflict with experiments that found
$G_R$ was proportional to $\rho_e$ \cite{vanMelick03,bartczak05}.
For the specific polymers studied, the entanglement density
of mixtures is a weighted average of pure systems.
Thus these systems also obey stress superposition.
It would be interesting to perform similar experiments on
polymer mixtures whose entanglement density was a nonlinear
function of weight fraction.

The relation between $\rho_e$ and $f$ is nonlinear for
the model polymers considered here, yet plots in a
previous paper indicated $G_R$ was proportional to $\rho_e$ 
\cite{hoy06}.
While this is approximately true for monodisperse systems,
the results for mixtures were misinterpreted in this earlier paper.
A factor of $(1-f)$ was inadvertently excluded in the calculation \cite{foot6}
of $\rho_e$ from Primitive Path Analysis \cite{everaers04}.
When $\rho_e$ is evaluated correctly, the hardening modulus is no longer
proportional to $\rho_e$.
However, we do find that the values of $G_r$ from Ref. \cite{hoy06}
satisfy stress superposition.
Indeed, the superposition principle explains the success of the previous
collapse.
As noted in Ref. \cite{hoy06}, the incorrectly determined value of $\rho_e$
was proportional to $(1-f)$ and so the collapse implied $G_R \propto (1-f)$
as expected from superposition.
The correct $\rho_e$ is nearly proportional to $(1-f)^2$
and not linearly related to $G_R$.

To understand why the contribution of chains to stress is additive,
we examined the changes in molecular conformations during shear.
As in Ref.\ \cite{hoy07}, the orientation was quantified by
the effective stretch $\lambda_{eff}$.
This is defined as the mean stretch of chains along the compression
direction; $\lambda_{eff} = \langle  R_z/R_z^0  \rangle$, where $R_z$ is
the rms $z$-component of the end-to-end vector of a chain and $R_z^0$ the
value before strain.
Panel (c) shows $\lambda_{eff}(\lambda)$ for short and long chains in the pure and mixed states for the same systems shown in panels (a-b).
Remarkably, $\lambda_{eff}^{long}$ and $\lambda_{eff}^{short}$ do not themselves depend on $f$.
In other words, the orientation of the long (short) chains is not dependent on whether they are in a pure system or mixed with a fairly high fraction of short (long) chains.
This $f$-independence of $\lambda_{eff}$ holds over the same broad
range of conditions as the stress superposition principle.

How can we understand the apparent $f$-independence of $\lambda_{eff}$?
The simplest qualitative picture is that interchain interactions
in the glass tend to enforce an affine deformation of the chains,
but that this is opposed by intrachain bonds.
Chains can not deform affinely at small scales because this would stretch
the stiff covalent bonds along the chains.
Instead, chains are observed to straighten over an effective persistence
length $l_p$ that increases with $g$ \cite{hoy06,vorselaars09b}.
This straightening is independent of chain length until the associated
stress becomes large enough to force relaxation of the chain along
its length.
For the systems considered here, the friction with neighboring chains
is large enough that significant relaxation does not occur until unentangled
chains are stretched to a substantial fraction of their fully
extended length.
Straightening chains while maintaining their connectivity requires 
local rearrangements of the surrounding monomers.
The number of rearrangements grows with the persistence length
and this was found to correlate directly with the strain hardening
at low temperatures \cite{mckechnie93,hoy07,hoy08,vorselaars09b}.

Note that entanglements do not enter directly in this picture.
Since entangled and unentangled chains behave in the same manner
at small $|g|$, the constraints on their rearrangements must
be determined by very local interactions with the surrounding matrix.
Entanglements
only become relevant when $g$ is so large that $l_p$ reaches a significant
fraction of the separation between entanglements \cite{hoy07,hoy08}.
In this limit entanglements become more effective than friction
from the matrix in
enforcing an affine displacement of the chain.
The associated tension in the chains grows as $l_p$ approaches
$N_e l_0$ and contributes significant energetic terms to the strain hardening
at large $|g|$ \cite{hasan93,chui99,li06,hoy08}.

It is worth considering under which conditions this mean field picture should
break down. 
For the reasons just described, the range of validity will be limited to
small strains in very densely entangled systems; fortunately, by this standard, few synthetic polymers are so densely entangled \cite{haward93}. 
The mean field picture should also break down for extremely short chains
where the high density of chain ends changes the density, friction and other 
properties significantly.
Stress superposition will be particularly sensitive to
changes in the relaxation rate of chains on the end-end scale, $\tau_{relax}^{-1}$, with $f$.
These are known to be significant for $T$ above $T_g$, where
$\tau_{relax}$ for short chains is increased when
they are mixed with longer chains, leading to greater orientation
at a given shear rate \cite{baljon95,ylitalo91}.
Our results show that such coupled-orientation effects
are small deep in the glassy state.

\section{Discussion and Conclusions}

The simulations presented here provide further evidence that
strain hardening in polymer glasses depends primarily on chain orientation
rather than entanglement.
They also provide new insight into the factors that control the degree
of orientation.
The local coupling to the glassy matrix enforces an affine deformation
of the chains ($\lambda=\lambda_{eff}$) at small $|g|$.
Hardening arises primarily from the plastic rearrangements needed
to maintain intrachain connectivity as chains orient and are straightened
over greater lengths.
As the associated stress on the chains grows, there is an increasing
tendency for them to relax along their length, leading to subaffine
deformation ($\lambda \neq \lambda_{eff}$).
The observation that the value of $\lambda_{eff}$
for a given chain length is independent of $f$
and that the stress due to plastic rearrangements obeys
the superposition principle, implies that chains relax
nearly independently.
The degree of relaxation depends on a competition between
the intrachain tension caused by alignment and friction
with the glassy matrix.
Entanglements only enter at very large $|g|$ where they
prevent relaxation of chains that are much longer than $N_e$.

To our knowledge, no microscopic theory that predicts the functional form
of $\lambda_{eff}$ in glasses has been published.
Constitutive models which employ a viscoelastic or viscoplastic
description of glassy strain hardening
(e.\ g.\ Refs.\ \cite{anand03,caruthers04,dupaix07})
typically decompose $\bar\lambda$ into
rubber-elastic and plastic parts
(or use other internal state variables), but do not explicitly account
for $\lambda \neq \lambda_{eff}$ or the $N$-dependence of nonaffine
relaxation \cite{foot8}.
It would be interesting to see if a first principles theory like that of  Ref.\ \cite{chen09} could be generalized to predict $\lambda_{eff}(\lambda)$.
The qualitative ingredients that must enter such a microscopic theory are
evident from the results shown here and in Refs. \cite{hoy07,hoy08,vorselaars09b}.
Simply put, $\lambda_{eff}$ falls behind $\lambda$ when the stress in the material is insufficient to  further affinely orient the average chain.

Fortuitously, $\lambda_{eff}$ can now be accurately measured in 
scanning near-field optical microscopy experiments \cite{ube07} which can discriminate values for different chain lengths in a bidisperse system.
Ref.\ \cite{ube07} showed $\lambda_{eff}$ falls behind $\lambda$ for entangled chains deformed slightly above $T_g$; similar studies on a variety of chain lengths below $T_g$ would be of great interest.
These might prove that $\lambda_{eff}$ is one of the ``mesoscale'' quantities \cite{barrat07} whose understanding will improve modeling of the deformation of amorphous materials.  
Other recent experiments \cite{hintermeyer08} have shown that segmental dynamics and larger-scale polymer dynamics are nontrivially coupled below $T_g$.
Further developments along these lines would be welcome, as would extension of the recently improved understanding of the relation \cite{grassia09} between linear viscoelastic behavior and structural relaxation to the nonlinear regime.

\acknowledgements
Kenneth S. Schweizer provided the original concept of $\lambda_{eff}$.
Daniel J.\ Read provided extensive, helpful correspondence regarding mixtures, and
Edward J.\ Kramer, Hugh R.\ Brown and Kang Chen provided helpful discussions.
All simulations were performed using the LAMMPS MD code \cite{lammpslink}.
Gary S. Grest provided some equilibrated melt states.
This material is based upon work supported by the National Science
Foundation under Grant No.\ DMR-0454947 and through the MRSEC
Program under Award No.\ DMR05-20415.


\begin{thebibliography}{50}
\expandafter\ifx\csname natexlab\endcsname\relax\def\natexlab#1{#1}\fi
\expandafter\ifx\csname bibnamefont\endcsname\relax
  \def\bibnamefont#1{#1}\fi
\expandafter\ifx\csname bibfnamefont\endcsname\relax
  \def\bibfnamefont#1{#1}\fi
\expandafter\ifx\csname citenamefont\endcsname\relax
  \def\citenamefont#1{#1}\fi
\expandafter\ifx\csname url\endcsname\relax
  \def\url#1{\texttt{#1}}\fi
\expandafter\ifx\csname urlprefix\endcsname\relax\def\urlprefix{URL }\fi
\providecommand{\bibinfo}[2]{#2}
\providecommand{\eprint}[2][]{\url{#2}}

\bibitem[{\citenamefont{Haward and Thackray}(1968)}]{haward68}
\bibinfo{author}{\bibfnamefont{R.~N.} \bibnamefont{Haward}} \bibnamefont{and}
  \bibinfo{author}{\bibfnamefont{G.}~\bibnamefont{Thackray}},
  \bibinfo{journal}{Proc. Roy. Soc. Lond.} \textbf{\bibinfo{volume}{302}},
  \bibinfo{pages}{453} (\bibinfo{year}{1968}).

\bibitem[{\citenamefont{Arruda and Boyce}(1993)}]{arruda93b}
\bibinfo{author}{\bibfnamefont{E.~M.} \bibnamefont{Arruda}} \bibnamefont{and}
  \bibinfo{author}{\bibfnamefont{M.~C.} \bibnamefont{Boyce}},
  \bibinfo{journal}{Int. J. Plast.} \textbf{\bibinfo{volume}{9}},
  \bibinfo{pages}{697} (\bibinfo{year}{1993}).

\bibitem[{foo({\natexlab{a}})}]{foot3}
\bibinfo{note}{Hereon tensor notation for $\bar\lambda$ is dropped for
  convenience, but $\lambda$ and $\lambda_{eff}$ remain tensorial.}

\bibitem[{\citenamefont{Kramer}(2005)}]{kramer05}
\bibinfo{author}{\bibfnamefont{E.~J.} \bibnamefont{Kramer}},
  \bibinfo{journal}{J. Polym. Sci. Part B - Polym. Phys.}
  \textbf{\bibinfo{volume}{43}}, \bibinfo{pages}{3369} (\bibinfo{year}{2005}).

\bibitem[{\citenamefont{van Melick et~al.}(2003)\citenamefont{van Melick,
  Govaert, and Meijer}}]{vanMelick03}
\bibinfo{author}{\bibfnamefont{H.~G.~H.} \bibnamefont{van Melick}},
  \bibinfo{author}{\bibfnamefont{L.~E.} \bibnamefont{Govaert}},
  \bibnamefont{and} \bibinfo{author}{\bibfnamefont{H.~E.~H.}
  \bibnamefont{Meijer}}, \bibinfo{journal}{Polymer}
  \textbf{\bibinfo{volume}{44}}, \bibinfo{pages}{2493} (\bibinfo{year}{2003}).

\bibitem[{\citenamefont{Dupaix and Boyce}(2005)}]{dupaix05}
\bibinfo{author}{\bibfnamefont{R.~B.} \bibnamefont{Dupaix}} \bibnamefont{and}
  \bibinfo{author}{\bibfnamefont{M.~C.} \bibnamefont{Boyce}},
  \bibinfo{journal}{Polymer} \textbf{\bibinfo{volume}{46}},
  \bibinfo{pages}{4827} (\bibinfo{year}{2005}).

\bibitem[{\citenamefont{Wendlandt et~al.}(2005)\citenamefont{Wendlandt,
  Tervoort, and Suter}}]{wendlandt05}
\bibinfo{author}{\bibfnamefont{M.}~\bibnamefont{Wendlandt}},
  \bibinfo{author}{\bibfnamefont{T.~A.} \bibnamefont{Tervoort}},
  \bibnamefont{and} \bibinfo{author}{\bibfnamefont{U.~W.} \bibnamefont{Suter}},
  \bibinfo{journal}{Polymer} \textbf{\bibinfo{volume}{46}},
  \bibinfo{pages}{11786} (\bibinfo{year}{2005}).
\bibinfo{author}{\bibfnamefont{M.}~\bibnamefont{Wendlandt}},
  \bibinfo{author}{\bibfnamefont{T.~A.} \bibnamefont{Tervoort}},
  \bibinfo{author}{\bibfnamefont{J.~D.} \bibnamefont{van Beek}},
  \bibnamefont{and} \bibinfo{author}{\bibfnamefont{U.~W.} \bibnamefont{Suter}},
  \bibinfo{journal}{J. Mech. Phys. Solids} \textbf{\bibinfo{volume}{54}},
  \bibinfo{pages}{589} (\bibinfo{year}{2006}).

\bibitem[{\citenamefont{Hine et~al.}(2007)\citenamefont{Hine, Duckett, and
  Read}}]{hine07}
\bibinfo{author}{\bibfnamefont{P.~J.} \bibnamefont{Hine}},
  \bibinfo{author}{\bibfnamefont{A.}~\bibnamefont{Duckett}}, \bibnamefont{and}
  \bibinfo{author}{\bibfnamefont{D.~J.} \bibnamefont{Read}},
  \bibinfo{journal}{Macromolecules} \textbf{\bibinfo{volume}{40}},
  \bibinfo{pages}{2782} (\bibinfo{year}{2007}).

\bibitem[{\citenamefont{Casas et~al.}(2008)\citenamefont{Casas,
  {Alba-Simionesco}, Montes, and Lequeux}}]{casas08}
\bibinfo{author}{\bibfnamefont{F.}~\bibnamefont{Casas}},
  \bibinfo{author}{\bibfnamefont{C.}~\bibnamefont{{Alba-Simionesco}}},
  \bibinfo{author}{\bibfnamefont{H.}~\bibnamefont{Montes}}, \bibnamefont{and}
  \bibinfo{author}{\bibfnamefont{F.}~\bibnamefont{Lequeux}},
  \bibinfo{journal}{Macromolecules} \textbf{\bibinfo{volume}{41}},
  \bibinfo{pages}{860} (\bibinfo{year}{2008}).

\bibitem[{\citenamefont{Govaert et~al.}(2008)\citenamefont{Govaert, Engels,
  Wendlandt, T.~A, and Suter}}]{govaert08}
\bibinfo{author}{\bibfnamefont{L.~E.} \bibnamefont{Govaert}},
  \bibinfo{author}{\bibfnamefont{T.~A.~P.} \bibnamefont{Engels}},
  \bibinfo{author}{\bibfnamefont{M.}~\bibnamefont{Wendlandt}},
  \bibinfo{author}{\bibfnamefont{T.}~\bibnamefont{T.~A}}, \bibnamefont{and}
  \bibinfo{author}{\bibfnamefont{U.~W.} \bibnamefont{Suter}},
  \bibinfo{journal}{J. Polym. Sci. Part B - Polym. Phys.}
  \textbf{\bibinfo{volume}{46}}, \bibinfo{pages}{2475} (\bibinfo{year}{2008}).

\bibitem[{\citenamefont{Hoy and Robbins}(2006)}]{hoy06}
\bibinfo{author}{\bibfnamefont{R.~S.} \bibnamefont{Hoy}} \bibnamefont{and}
  \bibinfo{author}{\bibfnamefont{M.~O.} \bibnamefont{Robbins}},
  \bibinfo{journal}{J. Polym. Sci. Part B - Polymer Phys.}
  \textbf{\bibinfo{volume}{44}}, \bibinfo{pages}{3487} (\bibinfo{year}{2006}).

\bibitem[{\citenamefont{Hoy and Robbins}(2007)}]{hoy07}
\bibinfo{author}{\bibfnamefont{R.~S.} \bibnamefont{Hoy}} \bibnamefont{and}
  \bibinfo{author}{\bibfnamefont{M.~O.} \bibnamefont{Robbins}},
  \bibinfo{journal}{Phys. Rev. Lett.} \textbf{\bibinfo{volume}{99}},
  \bibinfo{pages}{117801} (\bibinfo{year}{2007}).

\bibitem[{\citenamefont{Hoy and Robbins}(2008)}]{hoy08}
\bibinfo{author}{\bibfnamefont{R.~S.} \bibnamefont{Hoy}} \bibnamefont{and}
  \bibinfo{author}{\bibfnamefont{M.~O.} \bibnamefont{Robbins}},
  \bibinfo{journal}{Phys. Rev. E} \textbf{\bibinfo{volume}{77}},
  \bibinfo{pages}{031801} (\bibinfo{year}{2008}).

\bibitem[{\citenamefont{Lyulin et~al.}(2004)\citenamefont{Lyulin, Balabaev,
  Mazo, and Michels}}]{lyulin04}
\bibinfo{author}{\bibfnamefont{A.~V.} \bibnamefont{Lyulin}},
  \bibinfo{author}{\bibfnamefont{N.~K.} \bibnamefont{Balabaev}},
  \bibinfo{author}{\bibfnamefont{M.~A.} \bibnamefont{Mazo}}, \bibnamefont{and}
  \bibinfo{author}{\bibfnamefont{M.~A.~J.} \bibnamefont{Michels}},
  \bibinfo{journal}{Macromolecules} \textbf{\bibinfo{volume}{37}},
  \bibinfo{pages}{8785} (\bibinfo{year}{2004});
\bibinfo{author}{\bibfnamefont{A.~V.} \bibnamefont{Lyulin}},
  \bibinfo{author}{\bibfnamefont{B.}~\bibnamefont{Vorselaars}},
  \bibinfo{author}{\bibfnamefont{M.~A.} \bibnamefont{Mazo}},
  \bibinfo{author}{\bibfnamefont{N.~K.} \bibnamefont{Balabaev}},
  \bibnamefont{and} \bibinfo{author}{\bibfnamefont{M.~A.~J.}
  \bibnamefont{Michels}}, \bibinfo{journal}{Europhys. Lett,}
  \textbf{\bibinfo{volume}{71}}, \bibinfo{pages}{618} (\bibinfo{year}{2005}).

\bibitem[{\citenamefont{Robbins and Hoy}(2009)}]{robbins09}
\bibinfo{author}{\bibfnamefont{M.~O.} \bibnamefont{Robbins}} \bibnamefont{and}
  \bibinfo{author}{\bibfnamefont{R.~S.} \bibnamefont{Hoy}},
  \bibinfo{journal}{J. Polym. Sci. Part B - Polym. Phys.}
  \textbf{\bibinfo{volume}{47}}, \bibinfo{pages}{1406} (\bibinfo{year}{2009}).

\bibitem[{\citenamefont{Vorselaars
  et~al.}(2009{\natexlab{a}})\citenamefont{Vorselaars, Lyulin, and
  Michels}}]{vorselaars09}
\bibinfo{author}{\bibfnamefont{B.}~\bibnamefont{Vorselaars}},
  \bibinfo{author}{\bibfnamefont{A.~V.} \bibnamefont{Lyulin}},
  \bibnamefont{and} \bibinfo{author}{\bibfnamefont{M.~A.~J.}
  \bibnamefont{Michels}}, \bibinfo{journal}{J. Chem. Phys.}
  \textbf{\bibinfo{volume}{130}}, \bibinfo{pages}{074905}
  (\bibinfo{year}{2009}{\natexlab{a}}).

\bibitem[{\citenamefont{Vorselaars
  et~al.}(2009{\natexlab{b}})\citenamefont{Vorselaars, Lyulin, and
  Michels}}]{vorselaars09b}
\bibinfo{author}{\bibfnamefont{B.}~\bibnamefont{Vorselaars}},
  \bibinfo{author}{\bibfnamefont{A.~V.} \bibnamefont{Lyulin}},
  \bibnamefont{and} \bibinfo{author}{\bibfnamefont{M.~A.~J.}
  \bibnamefont{Michels}}, \bibinfo{journal}{Macromolecules}
  \textbf{\bibinfo{volume}{42}}, \bibinfo{pages}{5829}
  (\bibinfo{year}{2009}{\natexlab{b}}).

\bibitem[{\citenamefont{Chen and Schweizer}(2009)}]{chen09}
\bibinfo{author}{\bibfnamefont{K.}~\bibnamefont{Chen}} \bibnamefont{and}
  \bibinfo{author}{\bibfnamefont{K.~S.} \bibnamefont{Schweizer}},
  \bibinfo{journal}{Phys. Rev. Lett} \textbf{\bibinfo{volume}{102}},
  \bibinfo{pages}{038301} (\bibinfo{year}{2009}).


\bibitem[{\citenamefont{Argon}(1973)}]{argon73}
\bibinfo{author}{\bibfnamefont{A.~S.} \bibnamefont{Argon}},
  \bibinfo{journal}{Philos. Mag.} \textbf{\bibinfo{volume}{28}},
  \bibinfo{pages}{39} (\bibinfo{year}{1973}).

\bibitem[{\citenamefont{{G'Sell} and Souahi}(1997)}]{gsell97}
\bibinfo{author}{\bibfnamefont{C.}~\bibnamefont{{G'Sell}}} \bibnamefont{and}
  \bibinfo{author}{\bibfnamefont{A.}~\bibnamefont{Souahi}},
  \bibinfo{journal}{J. Eng. Materials Tech.} \textbf{\bibinfo{volume}{119}},
  \bibinfo{pages}{223} (\bibinfo{year}{1997}).

\bibitem[{\citenamefont{Mckechnie et~al.}(1993)\citenamefont{Mckechnie, Haward,
  Brown, and Clarke}}]{mckechnie93}
\bibinfo{author}{\bibfnamefont{J.~I.} \bibnamefont{Mckechnie}},
  \bibinfo{author}{\bibfnamefont{R.~N.} \bibnamefont{Haward}},
  \bibinfo{author}{\bibfnamefont{D.}~\bibnamefont{Brown}}, \bibnamefont{and}
  \bibinfo{author}{\bibfnamefont{J.~H.~R.} \bibnamefont{Clarke}},
  \bibinfo{journal}{Macromolecules} \textbf{\bibinfo{volume}{26}},
  \bibinfo{pages}{198} (\bibinfo{year}{1993}).

\bibitem[{\citenamefont{Fetters et~al.}(1994)\citenamefont{Fetters, Lohse,
  Richter, Witten, and Zirkel}}]{fetters94}
\bibinfo{author}{\bibfnamefont{L.~J.} \bibnamefont{Fetters}},
  \bibinfo{author}{\bibfnamefont{D.~J.} \bibnamefont{Lohse}},
  \bibinfo{author}{\bibfnamefont{D.}~\bibnamefont{Richter}},
  \bibinfo{author}{\bibfnamefont{T.~A.} \bibnamefont{Witten}},
  \bibnamefont{and} \bibinfo{author}{\bibfnamefont{A.}~\bibnamefont{Zirkel}},
  \bibinfo{journal}{Macromolecules} \textbf{\bibinfo{volume}{27}},
  \bibinfo{pages}{4639} (\bibinfo{year}{1994}).

\bibitem[{\citenamefont{Kremer and Grest}(1990)}]{kremer90}
\bibinfo{author}{\bibfnamefont{K.}~\bibnamefont{Kremer}} \bibnamefont{and}
  \bibinfo{author}{\bibfnamefont{G.~S.} \bibnamefont{Grest}},
  \bibinfo{journal}{J. Chem. Phys.} \textbf{\bibinfo{volume}{92}},
  \bibinfo{pages}{5057} (\bibinfo{year}{1990}).

\bibitem[{\citenamefont{Auhl et~al.}(2003)\citenamefont{Auhl, Everarers, Grest,
  Kremer, and Plimpton}}]{auhl03}
\bibinfo{author}{\bibfnamefont{R.}~\bibnamefont{Auhl}},
  \bibinfo{author}{\bibfnamefont{R.}~\bibnamefont{Everarers}},
  \bibinfo{author}{\bibfnamefont{G.~S.} \bibnamefont{Grest}},
  \bibinfo{author}{\bibfnamefont{K.}~\bibnamefont{Kremer}}, \bibnamefont{and}
  \bibinfo{author}{\bibfnamefont{S.~J.} \bibnamefont{Plimpton}},
  \bibinfo{journal}{J. Chem. Phys.} \textbf{\bibinfo{volume}{119}},
  \bibinfo{pages}{12718} (\bibinfo{year}{2003}).

\bibitem[{\citenamefont{Rottler and Robbins}(2003)}]{rottler03c}
\bibinfo{author}{\bibfnamefont{J.}~\bibnamefont{Rottler}} \bibnamefont{and}
  \bibinfo{author}{\bibfnamefont{M.~O.} \bibnamefont{Robbins}},
  \bibinfo{journal}{Phys. Rev. E} \textbf{\bibinfo{volume}{68}},
  \bibinfo{pages}{011507} (\bibinfo{year}{2003}).

\bibitem[{\citenamefont{Warren and Rottler}(2007)}]{warren07}
\bibinfo{author}{\bibfnamefont{M.}~\bibnamefont{Warren}} \bibnamefont{and}
  \bibinfo{author}{\bibfnamefont{J.}~\bibnamefont{Rottler}},
  \bibinfo{journal}{Phys. Rev. E} \textbf{\bibinfo{volume}{76}},
  \bibinfo{pages}{031802} (\bibinfo{year}{2007});
\bibinfo{author}{\bibfnamefont{R.~A.} \bibnamefont{Riggleman}},
  \bibinfo{author}{\bibfnamefont{K.~S.} \bibnamefont{Schweizer}},
  \bibnamefont{and} \bibinfo{author}{\bibfnamefont{J.~J.} \bibnamefont{{de
  Pablo}}}, \bibinfo{journal}{Macromolecules} \textbf{\bibinfo{volume}{41}},
  \bibinfo{pages}{4969} (\bibinfo{year}{2008}).

\bibitem[{\citenamefont{Everaers et~al.}(2004)\citenamefont{Everaers,
  Sukumaran, Grest, Svaneborg, Sivasubramanian, and Kremer}}]{everaers04}
\bibinfo{author}{\bibfnamefont{R.}~\bibnamefont{Everaers}},
  \bibinfo{author}{\bibfnamefont{S.~K.} \bibnamefont{Sukumaran}},
  \bibinfo{author}{\bibfnamefont{G.~S.} \bibnamefont{Grest}},
  \bibinfo{author}{\bibfnamefont{C.}~\bibnamefont{Svaneborg}},
  \bibinfo{author}{\bibfnamefont{A.}~\bibnamefont{Sivasubramanian}},
  \bibnamefont{and} \bibinfo{author}{\bibfnamefont{K.}~\bibnamefont{Kremer}},
  \bibinfo{journal}{Science} \textbf{\bibinfo{volume}{303}},
  \bibinfo{pages}{823} (\bibinfo{year}{2004}).

\bibitem[{foo({\natexlab{b}})}]{foot7}
\bibinfo{note}{A smaller system with the same parameters as in panel (a) was
  tested at $|\dot\epsilon| = 10^{-6}/\tau_{LJ}$ and gave equivalent results,
  but these were noisier due to finite size effects. Changing the rate by two
  orders of magnitude had only a modest effect on $\lambda_{eff}$, as expected
  from a logarithmic rate sensitivity.}

\bibitem[{\citenamefont{Hasan and Boyce}(1993)}]{hasan93}
\bibinfo{author}{\bibfnamefont{O.~A.} \bibnamefont{Hasan}} \bibnamefont{and}
  \bibinfo{author}{\bibfnamefont{M.~C.} \bibnamefont{Boyce}},
  \bibinfo{journal}{Polymer} \textbf{\bibinfo{volume}{34}},
  \bibinfo{pages}{5085} (\bibinfo{year}{1993}).

\bibitem[{\citenamefont{Bartczak}(2005)}]{bartczak05}
\bibinfo{author}{\bibfnamefont{Z.}~\bibnamefont{Bartczak}},
  \bibinfo{journal}{Macromolecules} \textbf{\bibinfo{volume}{38}},
  \bibinfo{pages}{7702} (\bibinfo{year}{2005}).

\bibitem[{foo({\natexlab{c}})}]{foot6}
\bibinfo{note}{We thank Alexei E.\ Likhtman and Daniel J.\ Read for pointing
  out an inconsistency that led to this discovery.}

\bibitem[{\citenamefont{Chui and Boyce}(1999)}]{chui99}
\bibinfo{author}{\bibfnamefont{C.}~\bibnamefont{Chui}} \bibnamefont{and}
  \bibinfo{author}{\bibfnamefont{M.~C.} \bibnamefont{Boyce}},
  \bibinfo{journal}{Macromolecules} \textbf{\bibinfo{volume}{32}},
  \bibinfo{pages}{3795} (\bibinfo{year}{1999}).

\bibitem[{\citenamefont{Li et~al.}(2006)\citenamefont{Li, Mulder, Vorselaars,
  Lyulin, and Michels}}]{li06}
\bibinfo{author}{\bibfnamefont{J.}~\bibnamefont{Li}},
  \bibinfo{author}{\bibfnamefont{T.}~\bibnamefont{Mulder}},
  \bibinfo{author}{\bibfnamefont{B.}~\bibnamefont{Vorselaars}},
  \bibinfo{author}{\bibfnamefont{A.~V.} \bibnamefont{Lyulin}},
  \bibnamefont{and} \bibinfo{author}{\bibfnamefont{M.~A. .~J.}
  \bibnamefont{Michels}}, \bibinfo{journal}{Macromolecules}
  \textbf{\bibinfo{volume}{39}}, \bibinfo{pages}{7774} (\bibinfo{year}{2006}).

\bibitem[{\citenamefont{Haward}(1993)}]{haward93}
\bibinfo{author}{\bibfnamefont{R.~N.} \bibnamefont{Haward}},
  \bibinfo{journal}{Macromolecules} \textbf{\bibinfo{volume}{26}},
  \bibinfo{pages}{5860} (\bibinfo{year}{1993}).

\bibitem[{\citenamefont{Baljon et~al.}(1995)\citenamefont{Baljon, Grest, and
  Witten}}]{baljon95}
\bibinfo{author}{\bibfnamefont{A.~R.~C.} \bibnamefont{Baljon}},
  \bibinfo{author}{\bibfnamefont{G.~S.} \bibnamefont{Grest}}, \bibnamefont{and}
  \bibinfo{author}{\bibfnamefont{T.~A.} \bibnamefont{Witten}},
  \bibinfo{journal}{Macromolecules} \textbf{\bibinfo{volume}{28}},
  \bibinfo{pages}{1835} (\bibinfo{year}{1995}).

\bibitem[{\citenamefont{Ylitalo et~al.}(1991)\citenamefont{Ylitalo, Kornfield,
  Fuller, and Pearson}}]{ylitalo91}
\bibinfo{author}{\bibfnamefont{C.~M.} \bibnamefont{Ylitalo}},
  \bibinfo{author}{\bibfnamefont{J.~A.} \bibnamefont{Kornfield}},
  \bibinfo{author}{\bibfnamefont{G.~G.} \bibnamefont{Fuller}},
  \bibnamefont{and} \bibinfo{author}{\bibfnamefont{D.~S.}
  \bibnamefont{Pearson}}, \bibinfo{journal}{Macromolecules}
  \textbf{\bibinfo{volume}{24}}, \bibinfo{pages}{749} (\bibinfo{year}{1991}).

\bibitem[{\citenamefont{Anand and Gurtin}(2003)}]{anand03}
\bibinfo{author}{\bibfnamefont{L.}~\bibnamefont{Anand}} \bibnamefont{and}
  \bibinfo{author}{\bibfnamefont{M.~E.} \bibnamefont{Gurtin}},
  \bibinfo{journal}{Int. J. Solids Structures} \textbf{\bibinfo{volume}{40}},
  \bibinfo{pages}{1465} (\bibinfo{year}{2003}).

\bibitem[{\citenamefont{Caruthers et~al.}(2004)\citenamefont{Caruthers, Adolf,
  Chambers, and Shrikhande}}]{caruthers04}
\bibinfo{author}{\bibfnamefont{J.~M.} \bibnamefont{Caruthers}},
  \bibinfo{author}{\bibfnamefont{D.~B.} \bibnamefont{Adolf}},
  \bibinfo{author}{\bibfnamefont{R.~S.} \bibnamefont{Chambers}},
  \bibnamefont{and}
  \bibinfo{author}{\bibfnamefont{P.}~\bibnamefont{Shrikhande}},
  \bibinfo{journal}{Polymer} \textbf{\bibinfo{volume}{45}},
  \bibinfo{pages}{4577} (\bibinfo{year}{2004}).

\bibitem[{\citenamefont{Dupaix and Boyce}(2007)}]{dupaix07}
\bibinfo{author}{\bibfnamefont{R.~B.} \bibnamefont{Dupaix}} \bibnamefont{and}
  \bibinfo{author}{\bibfnamefont{M.~C.} \bibnamefont{Boyce}},
  \bibinfo{journal}{Mech. Materials} \textbf{\bibinfo{volume}{39}},
  \bibinfo{pages}{39} (\bibinfo{year}{2007}).

\bibitem[{foo({\natexlab{d}})}]{foot8}
\bibinfo{note}{Ref.\ \cite{bergstrom98} considered $N$-dependent (reptational)
  relaxation of incompletely crosslinked elastomers above $T_g$ and it would be
  interesting to see if this approach could be adapted to glassy systems.}

\bibitem[{\citenamefont{Ube et~al.}(2007)\citenamefont{Ube, Aoki, Ito,
  Horinaka, and Takigawa}}]{ube07}
\bibinfo{author}{\bibfnamefont{T.}~\bibnamefont{Ube}},
  \bibinfo{author}{\bibfnamefont{H.}~\bibnamefont{Aoki}},
  \bibinfo{author}{\bibfnamefont{S.}~\bibnamefont{Ito}},
  \bibinfo{author}{\bibfnamefont{J.}~\bibnamefont{Horinaka}}, \bibnamefont{and}
  \bibinfo{author}{\bibfnamefont{T.}~\bibnamefont{Takigawa}},
  \bibinfo{journal}{Polymer} \textbf{\bibinfo{volume}{48}},
  \bibinfo{pages}{6221} (\bibinfo{year}{2007});
\bibinfo{author}{\bibfnamefont{T.}~\bibnamefont{Ube}},
  \bibinfo{author}{\bibfnamefont{H.}~\bibnamefont{Aoki}},
  \bibinfo{author}{\bibfnamefont{S.}~\bibnamefont{Ito}},
  \bibinfo{author}{\bibfnamefont{J.}~\bibnamefont{Horinaka}},
  \bibinfo{author}{\bibfnamefont{T.}~\bibnamefont{Takigawa}}, \bibnamefont{and}
  \bibinfo{author}{\bibfnamefont{T.}~\bibnamefont{Masuda}},
  \bibinfo{journal}{Polymer} \textbf{\bibinfo{volume}{50}},
  \bibinfo{pages}{3016} (\bibinfo{year}{2009}).

\bibitem[{\citenamefont{Barrat and {de Pablo}}(2007)}]{barrat07}
\bibinfo{author}{\bibfnamefont{J.}~\bibnamefont{Barrat}} \bibnamefont{and}
  \bibinfo{author}{\bibfnamefont{J.~J.} \bibnamefont{{de Pablo}}},
  \bibinfo{journal}{MRS Bull.} \textbf{\bibinfo{volume}{32}},
  \bibinfo{pages}{941} (\bibinfo{year}{2007}).

\bibitem[{\citenamefont{Hintermeyer et~al.}(2008)\citenamefont{Hintermeyer,
  Herrmann, Kahlau, Goiceanu, and R{\"o}ssler}}]{hintermeyer08}
\bibinfo{author}{\bibfnamefont{J.}~\bibnamefont{Hintermeyer}},
  \bibinfo{author}{\bibfnamefont{A.}~\bibnamefont{Herrmann}},
  \bibinfo{author}{\bibfnamefont{R.}~\bibnamefont{Kahlau}},
  \bibinfo{author}{\bibfnamefont{C.}~\bibnamefont{Goiceanu}}, \bibnamefont{and}
  \bibinfo{author}{\bibfnamefont{E.~A.} \bibnamefont{R{\"o}ssler}},
  \bibinfo{journal}{Macromolecules} \textbf{\bibinfo{volume}{41}},
  \bibinfo{pages}{9335} (\bibinfo{year}{2008}).

\bibitem[{\citenamefont{Grassia and {D'Amore}}(2009)}]{grassia09}
\bibinfo{author}{\bibfnamefont{L.}~\bibnamefont{Grassia}} \bibnamefont{and}
  \bibinfo{author}{\bibfnamefont{A.}~\bibnamefont{{D'Amore}}},
  \bibinfo{journal}{J. Polym. Sci. Part B - Polym. Phys.}
  \textbf{\bibinfo{volume}{47}}, \bibinfo{pages}{724} (\bibinfo{year}{2009}).

\bibitem[{lam()}]{lammpslink}
\bibinfo{note}{\lowercase{h}ttp://lammps.sandia.gov/}.

\bibitem[{\citenamefont{Bergstr{\"o}m and Boyce}(1998)}]{bergstrom98}
\bibinfo{author}{\bibfnamefont{J.~S.} \bibnamefont{Bergstr{\"o}m}}
  \bibnamefont{and} \bibinfo{author}{\bibfnamefont{M.~C.} \bibnamefont{Boyce}},
  \bibinfo{journal}{J. Mech. Phys. Solids} \textbf{\bibinfo{volume}{46}},
  \bibinfo{pages}{931} (\bibinfo{year}{1998}).

\end{thebibliography}
\end{document}